\documentclass[twocolumn, tighten]{aastex61}
\usepackage{amsmath,amstext}
\usepackage{hyperref}

\shorttitle{Distant Halo stars with HSC}
\shortauthors{Deason et al.}

\begin{document}
\title{To the Galactic Virial Radius with Hyper Suprime-Cam}

\correspondingauthor{Alis J. Deason}
\email{alis.j.deason@durham.ac.uk}

\author[0000-0001-6146-2645]{Alis J. Deason}
\affiliation{Institute for Computational Cosmology, Department of Physics, University of Durham, South Road, Durham DH1 3LE, UK}

\author{Vasily Belokurov}
\affiliation{Institute of Astronomy, University of Cambridge, Madingley Road, Cambridge CB3 0HA, UK}
\affiliation{Center for Computational Astrophysics, Flatiron Institute, 162 5th Avenue, New York, NY 10010, USA}

\author[0000-0003-2644-135X]{Sergey E. Koposov}
\affiliation{McWilliams Center for Cosmology, Department of Physics, Carnegie Mellon University, 5000 Forbes Avenue, Pittsburgh, PA 15213, USA}
\affiliation{Institute of Astronomy, University of Cambridge, Madingley Road, Cambridge CB3 0HA, UK}

\begin{abstract}
We exploit the exquisite, deep Hyper Suprime-Cam (HSC) imaging data to probe the Galactic halo out to 200 kpc. Using the $\sim 100$ square degree, multi-band photometry of the first HSC Wide survey data release, we identify blue horizontal branch (BHB) stars beyond 50 kpc in the halo. The presence of the Sagittarius (Sgr) stream in the HSC fields produces a notable excess of stars at the apocentre of the leading arm ($\sim 50-60$ kpc). For fields excluding Sgr, the BHB counts are consistent with a continuation of a $-4$ power-law from the inner halo. However, we find that the majority of the non-Sgr BHB stars beyond 50 kpc reside in one 27 square degree HSC field called ``VVDS''. Curiously, this field is located close to the Magellanic plane, and we hypothesize that the excess of stars between 50 and 200 kpc could be associated with distant Magellanic debris. Indeed, without the VVDS, there are very few BHBs in the remaining portions of the Galaxy probed by the HSC. Accordingly, this scarcity of tracers is consistent with a significant decline in stellar density beyond 50 kpc, with a power-law of $-4$ or steeper. 
\end{abstract}

\keywords{Galaxy: halo -- Galaxy: structure -- Galaxy: stellar content -- galaxies: Magellanic Clouds}

\section{Introduction}
The ``Pale Blue Dot'' image is a tiny portion of one of the large
array of photographs sent back to Earth by the Voyager 1 mission. This
unassuming postcard from the outer Solar System provided a striking
perspective on our place in the local Universe; a humbling and
thought-provoking point of view. On Galactic scales, there is no direct
analog to launching a physical probe ``as far as the fuel takes you''. There is
no simple way to see what is out there or to cast a glance
back. However, there exists a small number of stars that travel to the
periphery (and perhaps even one step beyond) of the Milky Way which, if detected, can be used as tracers of its extent. In
particular, in the past decade, there have been several attempts to
expand on the inventory of these outer halo denizens \citep[see
  e.g.][]{Brown2010, deason12,Bochanski2014,Huxor2015, Cohen2017}.

These lonely Galactic scouts are precious as there is currently a
noticeable undersupply of information on the space beyond 100 kpc from
the Sun. Only 3 of $\sim$ the 150 Milky Way globular clusters are
currently located this far
\citep[][]{harris10,crater,laevens14}. While almost a half of the
Galaxy's dwarf satellites lie out there
\citep[][]{alan_dwarfs,crater2,aqu2}, this distant population is still
minuscule in terms of absolute numbers (N$<30$). Note that this mismatch
in Galactocentric distance preferences between the two groups of
satellites is the result of i) distinct creation and accretion
pathways, and ii) large differences in central densities and hence their
resistance to Galactic tides. Rare as they are, the dwarfs
nonetheless dictate that the gravitational influence of our Galaxy
\citep[see][for discussion]{shull14} must stretch beyond 100 kpc and
perhaps as far as 300 kpc. The existence of a large halo around the
Milky Way is backed up by observations of a hot gaseous corona
\citep[e.g.][]{tumlinson11}. Yet, it is the study of the stellar
component of the halo that appears the most promising: as tracers, stars
are more numerous (compared to the satellites) while still providing
secure individual distances (unlike gas).

If followed-up kinematically, swarms of these ``voyager'' stellar
particles can be used to put constraints on the total mass budget of
the Galaxy. Note, however, that no matter what style mass estimator is
used \citep[see e.g.][]{battaglia, watkins, anmass1, anmass2}, one
necessary assumption is the relaxedness of the stellar
halo. Additionally, the pre-requisite ingredient in the mass inference
is the size of the stellar cloud \citep[see][for
  discussion]{dehnen}. Unfortunately, the condition of complete
virialisation of the distant stellar halo is unattainable as the mixing times beyond a few tens of kpc from the Sun are
prohibitively long. To make matters worse, consensus is lacking as to
the details of the stellar density distribution in the periphery of
the Galaxy. The first attempt to secure accurate measurements of the
stellar halo beyond 50 kpc is recorded by \citet{deason14}. They used
stacked SDSS imaging - the deepest wide multi-band photometry
available at the time - to disentangle the behavior of faint Blue
Horizontal Branch (BHB), Blue Straggler (BS) and Wide Dwarf (WD) stars
in the presence of QSO contamination. This experiment revealed a
dramatic drop in the BHB density beyond 50 kpc: According to
\citet{deason14}, the number counts of this particular tracer are
consistent with power law indexes as steep as $-6$ or even
$-8$. Subsequently, several other groups endeavored to probe the outer
halo, each with a different sort of stellar tracer. For example,
\citet{Xue2015} used spectroscopically selected K giants delivered as
part of the SDSS survey, \citet{Slater2016} combined proprietary
narrow-band imaging with the SDSS broad-band photometry to single out
the Red Giant Branch (RGB) population, and, finally, \citet{Cohen2016, Cohen2017}
relied on the RR Lyrae stars from the Palomar Transient Facility (PTF)
archives. Curiously, only the measurement by \citet{Xue2015} is in
broad agreement with the results of \citet{deason14}. Both
\citet{Slater2016} and \citet{Cohen2016, Cohen2017} report significantly
shallower density profiles for the stellar halo between 50 and 100
kpc.

There are many possible explanations for this disagreement. First and
foremost, the different methods employed to select tracers are subject
to distinct completeness and contamination functions. While none of
the approaches mentioned above are completely free of contaminants, RR
Lyrae and spectroscopically selected giants are perhaps the cleanest
of the four. However, it is not the overall levels of completeness
and/or contamination that determine the success of the experiment. It
is the knowledge of their evolution across the dataset that is vital:
A model for completeness and contamination must be incorporated in the
likelihood model. On the other hand, it is not impossible that all
four of the measurements above are actually correct. Significant
population gradients could be present in the outer halo, amplified by
differences in the sky coverage and masked by the low-number
statistics. In fact, the question of the footprint and the depth of
each survey is especially important as the stellar component of the
halo is expected to be highly anisotropic \citep[see e.g.][]{BJ2005,
  Helmi2011, Libeskind2011}. While we have reasons to believe that the
outer halo is dominated by sub-structure, very few actual detections
have been recorded so far. Beyond 50 kpc, on scales larger than
several degrees, currently only two confirmed stellar debris
agglomerations are known. These are the giant stream from the Sgr
dwarf \citep[see][]{Newberg_BHB,Drake2013,belokurov14, Sesar2017} and
the Pisces Over-density
\citep[see][]{Sesar2007,Watkins_Pisces,Nie2015}.

And then there is the elephant in the room: The Magellanic
Clouds. Nearly half a century ago, a giant stream of neutral hydrogen
was discovered, attached to the LMC and the SMC on the sky
\citep[][]{WW1972,Mathewson1974}. Today, mapped in exquisite detail
\citep[see e.g.][]{Putman2003,Nidever2008,Nidever2010}, the Magellanic
Stream (MS) is yet to have a comprehensive model, but is generally
explained as a result of tidal interaction between the Clouds and the
Milky Way \citep[see][]{Besla2007, Besla2010, Diaz2011,
  Diaz2012}. Note, however, that a ram-pressure origin of the MS has
also been explored \citep[see e.g.][]{moore94, mastropietro05, Hammer2015, salem15}. What distinguishes
the tidal and the ram-pressure scenarios is that the former always
produces a stellar counter-part to both the Leading and Trailing
portions of the Stream. Thus the stellar debris from the Clouds should
be polluting the Galaxy across a wide range of distances; from tens
of kpc from the Sun out to the virial radius. Unfortunately, the
existence of the stellar component of the MS has not yet been
unambiguously confirmed. Nonetheless, to date, several groups have
announced detections of diffuse sprays of stellar material at large
distances from the LMC \citep[][]{Majewski1999,
  Munoz2006,Majewski2009,belokurov16}. Most recently,
the Gaia Data Release 1 photometry has been used to show that,
quite possibly, the stellar counterpart to the gaseous MS may be as
extended and as prominent as the models professed
\citep[see][]{belokurov17,deason17}.

Even if --- as both theory and observations indicate --- the outer
environs of the Galaxy are not fully virialised,
the shape of the stellar halo's radial density profile can inform our understanding of the accretion history of the Milky
Way. Using a suite of semi-analytic simulations by \citet{BJ2005},
\citet{Deason2013} demonstrated that the steepness of the outer
stellar halo is linked to the satellite accretion rate. For example,
if the stellar halo assembles most of its mass in an early short burst
of activity brought about by a significant merger, its outer profile
at z=0 would appear much steeper compared to a steadily growing
halo. This idea has since been confirmed by \citet{Pillepich2014}
using galaxies in the Illustris simulation suite. Compared to the
simulated galaxies, the fast density fall-off in the Milky Way appears
to indicate an early-peaked, and subsequently quiescent accretion
history. Moreover, as both \citet{Deason2016_eating} and
\citet{Amorisco2017} elucidate, this fossilized appearance of the
Galactic stellar halo is a transient phenomenon; it will
transform into a much younger looking and more metal-rich object as
soon as the debris from the Sgr dwarf and the Magellanic Clouds have
been fully digested.

In this work, we strive to clarify the behavior of the Galactic
stellar halo at distances beyond 50 kpc using the freshly released
imaging data from the Hyper Suprime-Cam mounted on the Subaru 8m
telescope. We take advantage of the unprecedented quality and depth of
the HSC multi-band photometry covering in excess of 100 square degrees
along multiple sight-lines through the Milky Way's halo to select BHB
stars with distances as large as $\sim200$ kpc. We discuss the
properties of the HSC dataset (such as the completeness and
contamination) and the details of the star-galaxy separation in
Section~\ref{sec:hsc}. Our modeling procedure is explained in
Section~\ref{sec:model}. The resulting BHB density profile is
presented in Section~\ref{sec:counts}. Curiously, we find signs of the
distant stellar debris possibly associated with the Magellanic Clouds'
in-fall --- this is discussed in Section~\ref{sec:magellanic}. Finally,
we put our measurements of the Milky Way's outer stellar halo into
context in Section~\ref{sec:final}.

\begin{figure*}
    \centering
    \includegraphics[width=16cm, height=5.33cm]{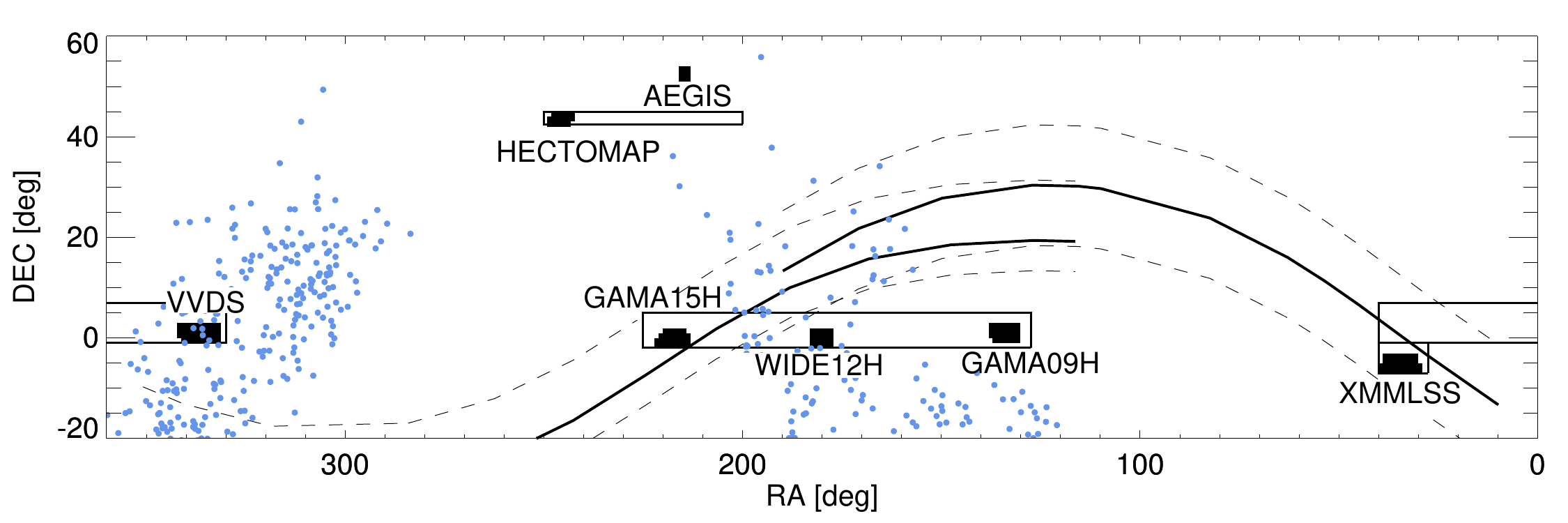}
    \caption{The sky coverage of the Hyper Suprime-Cam Wide survey in equatorial coordinates. The dark shaded regions show the coverage of the first data release, and we give the associated field names. The open boxes indicate the planned coverage of the entire Wide survey. The solid and dashed lines show the approximate track of the Sagittarius stream. Two of the fields, XMMLSS and GAMA15H, overlap with the leading and trailing arms of the stream. The blue points indicate the position on the sky of potential Magellanic debris between $ 50 < D/\mathrm{kpc} < 100$ from the \cite{Diaz2012} models (see Section \ref{sec:magellanic}).}
    \label{fig:fields}
\end{figure*}

\section{Hyper Suprime-Cam Photometry}
\label{sec:hsc}

Hyper Suprime-Cam (HSC; \citealt{miyazaki12}) is an optical imaging camera installed on the 8-m Subaru telescope. The camera has a very large field-of-view (1.5 deg diameter) and has five broad band filters, $g, r, i, z, y$. A large imaging survey is being conducted on HSC consisting of three layers: Wide, Deep, and UltraDeep. The Wide survey will cover 1,400 square degrees of the sky in all five broad band filters down to $i \sim 26$ at $5\sigma$ for point sources. In this work, we make use of the first data release of the Wide survey described in \citet{aihara17}. This release includes the data from the first 1.7 years of observations, and covers approximately 108 deg$^2$. The distribution of the Wide HSC fields in equatorial coordinates are shown in Fig. \ref{fig:fields}. In this figure, we also show the approximate track of the Sagittarius (Sgr) stream over the sky, and the predicted debris from models of Magellanic cloud disruption (see Section \ref{sec:magellanic}). Two of the fields (GAMA15H and XMMLSS) lie directly on the Sgr stream, and we will show in Section \ref{sec:counts} how this substructure affects our results.

\subsection{Star-galaxy separation}
\label{sec:star_gal}
\begin{figure}
    \centering
    \includegraphics[width=8.5cm, height=5.67cm]{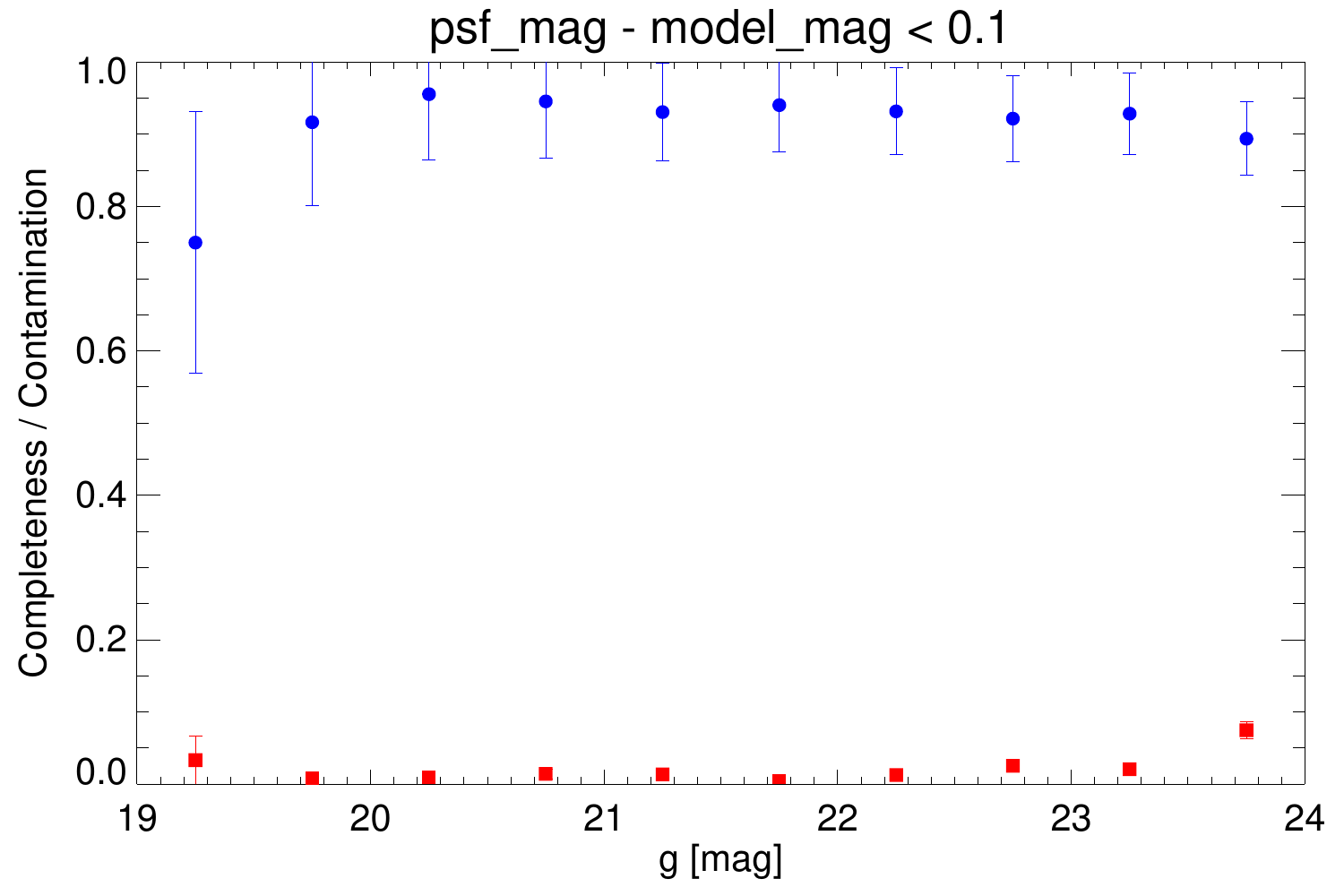}
    \caption{The completeness (blue filled circles) and contamination (red filled squares) of stellar sources as a function of g-band magnitude. We select stars by imposing a cut on the \textsc{psf\_mag - model\_mag} $< 0.1$ for the $griz$ photometry, and we use the overlapping \textit{HST}/ACS  catalog in COSMOS \citep{leauthaud07} as the star-galaxy separation ``truth'' table. For the magnitude range considered in this work, $19 <g < 22$ there is minimal contamination from galaxies and the completeness is $\gtrsim 90\%$.}
    \label{fig:comp_contam}
\end{figure}
In this work, we exploit the deep HSC photometry to identify stars in the distant halo of the Galaxy. To select point sources, we impose the cut: \textsc{psf\_mag -- model\_mag} $< 0.1$ for $g, r, i, z$ (see \citealt{aihara17}). We compute the completeness and contamination of our point source selection using the overlapping \textit{HST}/ACS  catalog in COSMOS \citep{leauthaud07}. Here, we assume that the \textit{HST}/ACS star/galaxy separation is the ``truth'', and compare with our selection of stars from HSC photometry. In Fig. \ref{fig:comp_contam} we show the resulting completeness and contamination of stellar sources as a function of magnitude. For the magnitude range considered in this work, $19 <g < 22$, there is minimal contamination from galaxies and the completeness is $\gtrsim 90\%$. Note that the saturation limit for the HSC photometry is $g \sim 18$, so we only consider stars fainter than this magnitude limit.

\subsection{Selection of Blue Horizontal Branch Stars}
\label{sec:selection}

\begin{figure*}
    \centering
    \includegraphics[width=16cm, height=16cm]{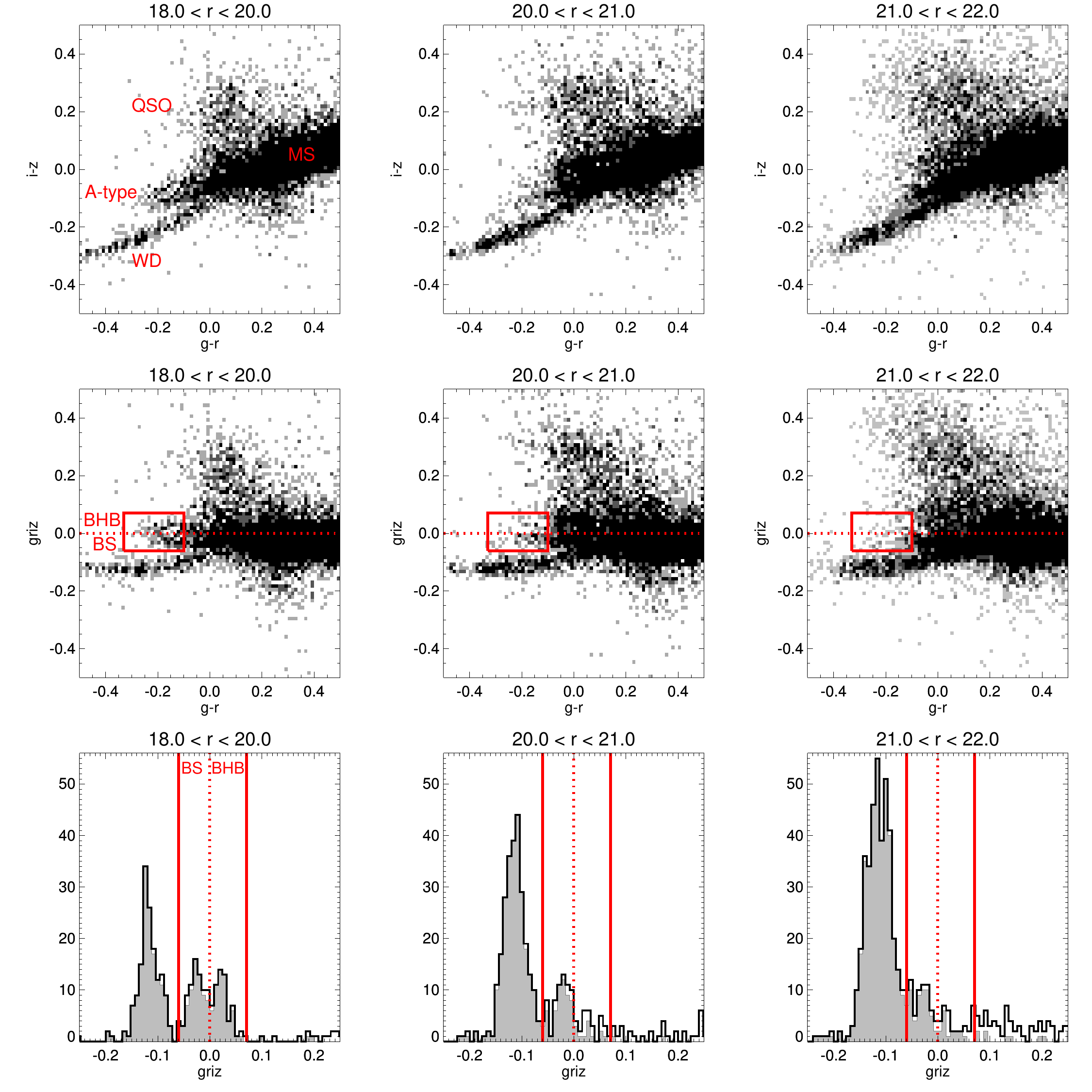}
    \caption{Selection of A-type stars in $g, r, i, z$ space. Each column shows different ranges of $r$-band magnitudes, becoming progressively fainter from left to right. The top panels show the distribution of stars in $i-z$ vs. $g-r$. The narrow strip at $g-r \lesssim-0.2$ are WDs. The A-type stars lie just above the WDs, on the blue-side of the main sequence stars. The middle panels show the combination of $g, r, i, z$, which flattens the A-type star sequence: $griz = i-z-0.3(g-r)+0.035$. The red box indicates the approximate selection of A-type stars and the red dotted line indicates the mid-plane between BHB and BS stars. The black histograms in the bottom panels shows the distribution of $griz$ for $-0.33 < g-r < -0.1$. The filled gray histograms show the distributions with an additional cut to remove QSO contamination (see Fig. \ref{fig:griz_ri}). The solid red lines indicate the approximate boundary of the A-type stars. WDs dominate at $griz < -0.1$ and QSOs contribute at $griz > 0.1$. The A-type stars become more blurred in $griz$ at fainter magnitudes and the contamination by WDs and QSOs increases.}
    \label{fig:griz}
\end{figure*}

\begin{figure}
    \centering
    \includegraphics[width=8.5cm, height=12.75cm]{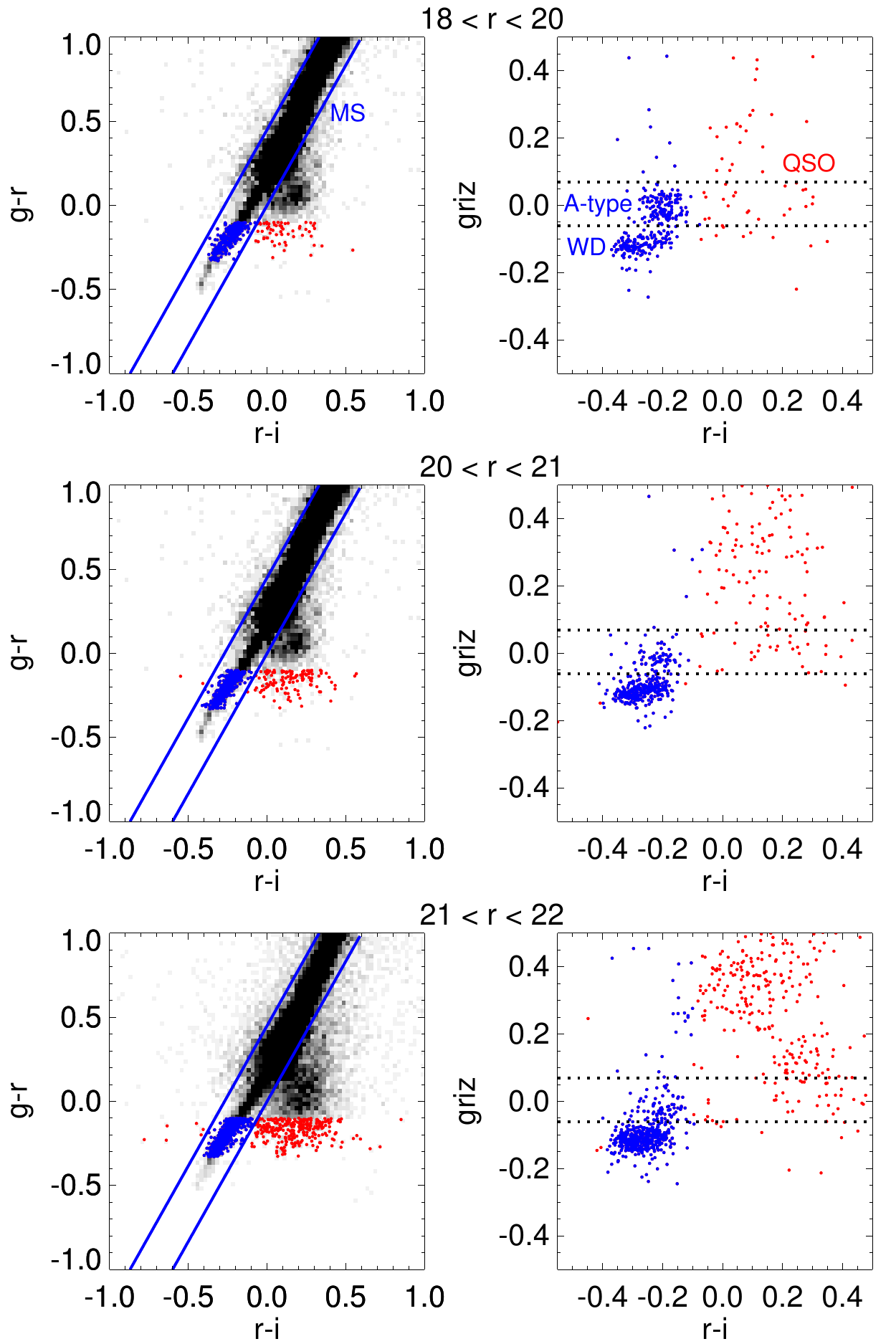}
    \caption{Further tuning of our A-type star selection to remove QSOs. The left hand panels show $g-r$ vs. $r-i$, and the right-hand panels show $griz$ vs. $r-i$. The thick blue lines in $g-r, r-i$ space bound the stellar sequence. The QSO contribution becomes apparent at redder $r-i$. The blue points indicate the stars selected with $-0.33 < g-r < -0.1$, and $0.6(g-r)+0.45 < r-i < 0.6(g-r)$, and the red points show stars that lie outside of the $g-r, r-i$ cut. This additional cut removes a significant number of QSOs (seen at $r- i > 0$) in the A-type star region of the $griz$ plane.}
    \label{fig:griz_ri}
\end{figure}

We aim to identify BHB stars from the deep HSC photometry. These old, metal-poor stars have bright absolute magnitudes ($M_g \sim 0.5$) and well-defined absolute magnitude calibrations, and thus are ideal tracers of the distant halo (e.g. \citealt{xue08, deason11, deason12}).  In the magnitude range under consideration in this work, $ 19 < g < 22$, BHBs trace from 50 kpc to 200 kpc in the Galactic halo -- i.e. out to the virial radius of the Galaxy!

Previous works have used multi-band photometry to select BHB stars \citep{lenz98, yanny00, deason11}, however, this selection largely relies on the  $u-g$ color to distinguish between BHBs and their contaminants. The $u-g$ color is a measurement of the near-UV flux excess, and can easily delineate A-type stars from white dwarfs (WDs) and QSOs. Moreover, with accurate enough photometry, the $u-g$ provides a subtle distinction between BHB stars and similar temperature, but higher surface gravity blue straggler (BS) stars. For example, \cite{deason11} used $u, g, r$ photometry in SDSS to model the density profile of BHB and BS stars out to $D \sim 40$ kpc. Here, the BHB and BS populations comprise distinct, but overlapping sequences in $u-g, g-r$ color-color space, and the number counts of these stars can be modeled probabilistically. There is no $u$ band in the HSC photometry; however, recent work has shown that near-IR photometry can also be used, with comparable success, to differentiate BHB stars from WDs, QSOs and BSs (see e.g. \citealt{vickers12, belokurov16}). In this work, we adopt a similar approach using a combination of $g, r, i, z$ to tease out the BHB signal.

In the top panels of Fig. \ref{fig:griz} we show $i-z$ vs. $g-r$ for stellar sources in the HSC Wide fields. Here, each panel shows a different $r$ band magnitude range (going fainter from left to right). The main sequence stars are clear at $g-r > 0$. The structure leading off from the main sequence to bluer colors are A-type stars, and the thinner sequence extending below the A-type stars are WDs. QSOs appear ``cloud-like'' in this color-color space and permeate the stellar main sequence, especially at redder $g-r > 0$.  In the middle panels, we show the combination of $g, r, i, z$ which delineates the A-type star sequence. Here, $griz= i-z-0.3(g-r)+0.035$. The red box indicates the A-type stars, where BHBs have $griz > 0$ and BSs have $griz < 0$. The distinction between BHBs and BSs is clearer in the bottom panels where we show a histogram of the $griz$ distribution for $-0.33 < g-r < -0.1$. The WDs are prominent at $griz < -0.1$ and the BHBs and BSs occupy narrow distributions either side of $griz =0$. The QSOs are apparent at $griz > 0.1$, and become more significant at fainter magnitudes.

We can further limit the contamination from QSOs using the $r-i$ color. Figure \ref{fig:griz_ri} shows $g-r$ vs. $r-i$ (left panels) and $griz$ vs. $r-i$ (right panels). Here, each row shows a different magnitude range (fainter from top to bottom). In the left-hand panels, the blue lines indicate the stellar sequence, where $0.6(g-r)+0.45 < r-i < 0.6(g-r)$. The QSOs don't occupy the same narrow sequence and are dispersed over a larger range of $r-i$ at fixed $g-r$. The blue points show stars in the A-type color range ($-0.33 < g-r < -0.1$) that lie within the stellar sequence. The red points indicate objects in the same $g-r$ color range that lie outside of the narrow stellar sequence, which are likely QSOs. Note that even at relatively faint magnitudes, the A-type stars occupy a narrow sequence. In the right-hand panels we show $griz$ vs. $r-i$. Here, it is clear that in the approximate $griz$ range of A-type stars, there are a significant number of QSOs (at redder $r-i$). However, by applying the cut: $0.6(g-r)+0.45 < r-i < 0.6(g-r)$, the majority of QSO contamination is excluded.

In the following Section, we use the $griz$ color to estimate the number of BHB stars as a function of distance modulus, and thus measure the stellar halo density profile out to 200 kpc.

\section{Modeling}
\label{sec:model}

\begin{figure}
    \centering
    \includegraphics[width=8.5cm, height=6.8cm]{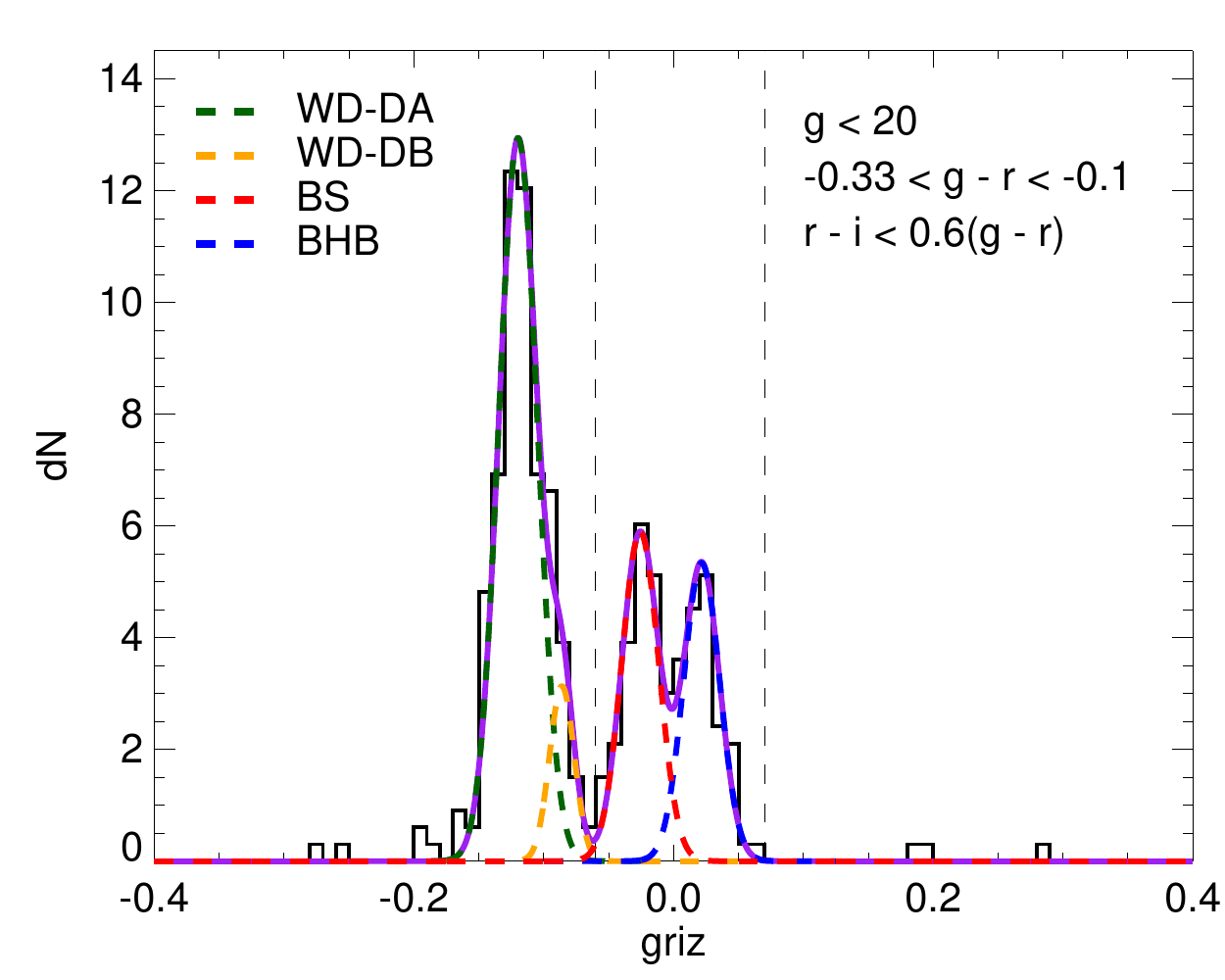}
    \caption{The Gaussian decomposition of relatively bright stars ($g < 20$) in the $griz$ plane.  We fit four Gaussian components, two to describe the WD contribution (the Hydrogen dominated DA-type and Helium dominated DB-type populations - green and yellow dashed lines), one for the blue stragglers (red dashed line) and one for the BHBs (blue dashed line). We use this decomposition to fix the centers and intrinsic widths of the Gaussian populations. Here, the photometric errors are very small (see Fig. \ref{fig:abm}) so we assume that the widths of these Gaussians are the \textit{intrinsic} widths of the populations. At fainter magnitudes the distributions will broaden with larger photometric errors.}
    \label{fig:cent_widths}
\end{figure}

\begin{figure}
    \centering
    \includegraphics[width=8.0cm, height=6.4cm]{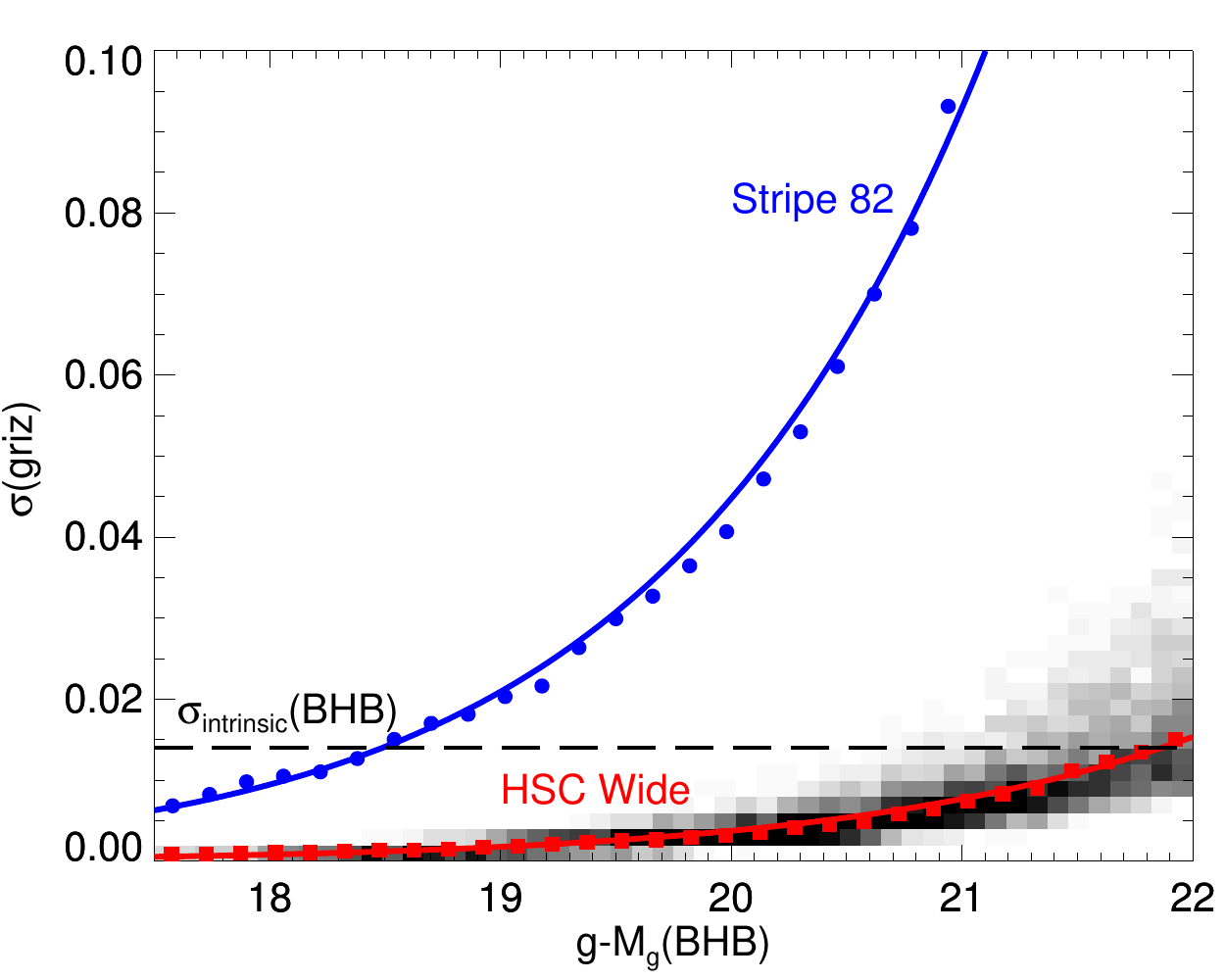}
    \caption{The uncertainty in $griz$ magnitude space as a function of BHB distance modulus. The red squares indicate the median values for the HSC photometry in bins of distance modulus and the solid red line shows a parametric fit. We use this fit to describe the increase in width of the $griz$ distributions (see Fig. \ref{fig:cent_widths}) with increasing distance modulus, e.g. $\sigma^2(\rm BHB) = \sigma^2_{\rm intrinsic}(BHB) + \sigma^2 (griz)$.  The dashed line indicates the \textit{intrinsic} width of the BHB distribution in $griz$ space. Finally, for comparison, we show the $griz$ uncertainty for the Stripe 82 photometry with the blue filled points. The HSC photometry is vastly superior in this magnitude range.}
    \label{fig:abm}
\end{figure}

The A-type stars, WDs and QSOs populate the $griz$ plane with distinct, but overlapping sequences. To help isolate BHB stars, we focus on objects with blue colors, $-0.33 < g-r < -0.1$ and remove significant QSO contamination by applying the cut: $0.6(g-r)+0.45 < r-i < 0.6(g-r)$. We assume WDs, BHBs and BSs have Gaussian distributions in $griz$. We fix the centres and \textit{intrinsic} widths\footnote{Note that it's likely that a systematic photometric uncertainty floor in the photometry also contributes to these intrinsic widths.}  of these Gaussians using relatively bright $g < 20$ stars, which have very small photometric errors (see Fig. \ref{fig:abm}). In Fig. \ref{fig:cent_widths} we show the Gaussian decomposition of relatively bright stars ($g < 20$) in the $griz$ plane. For WDs, we fit two components --- these are the H-dominated (DA-type, green dashed line), and He-dominated (DB-type, yellow dashed line) populations. The derived ratio between these two WD populations ($f_{\rm DA}=0.7$, $f_{\rm DB}=0.3$) is in good agreement with WD models (see Appendix A of \citealt{deason14}), and we fix this ratio for the remainder of the analysis. The BS and BHB distributions are shown with the red and blue dashed lines, respectively. At these bright magnitudes, the A-type stars have clearly distinct $griz$ distributions. The centres and intrinsic widths of the populations are given in Table \ref{tab:cent_widths}.

At fainter magnitudes, these populations will blur in $griz$ space, and, moreover, QSOs may start to leak into the A-type star $griz$ region. We model the variation of the Gaussian widths  with magnitude using the HSC photometric errors, e.g. $\sigma^2(\rm BHB) = \sigma^2_{\rm intrinsic}(BHB) + \sigma^2 (griz)$. Here, the intrinsic widths, calculated from the brightest stars (see Fig. \ref{fig:cent_widths}), are kept fixed. In Fig. \ref{fig:abm} we show the uncertainty in $griz$ magnitude space as a function of BHB distance modulus. Here, we use the absolute magnitude calibrations for BHB stars as a function of $g-r$ color from \cite{deason11}. Thus, for bins in BHB distance modulus, we have defined the centres and widths of the Gaussian populations, and we determine the amplitudes of the Gaussians using a a maximum likelihood analysis.

\begin{table}
\caption{The centres and intrinsic widths of the Gaussian $griz$ distributions. These values are kept fixed in our analysis.}
\label{tab:cent_widths}
\begin{center}
\begin{tabular}{lcc}
\hline
Type & $\langle griz \rangle$ & $\sigma_{\rm intrinsic} (griz)$\\
\hline
BHB & 0.022 & 0.014\\
BS &  -0.026 & 0.014 \\
WD-DA &  -0.120 & 0.015 \\
WD-DB & -0.086 & 0.010\\
\hline
\end{tabular}
\end{center}
\end{table}

\begin{figure*}
    \centering
    \includegraphics[width=15cm, height=5cm]{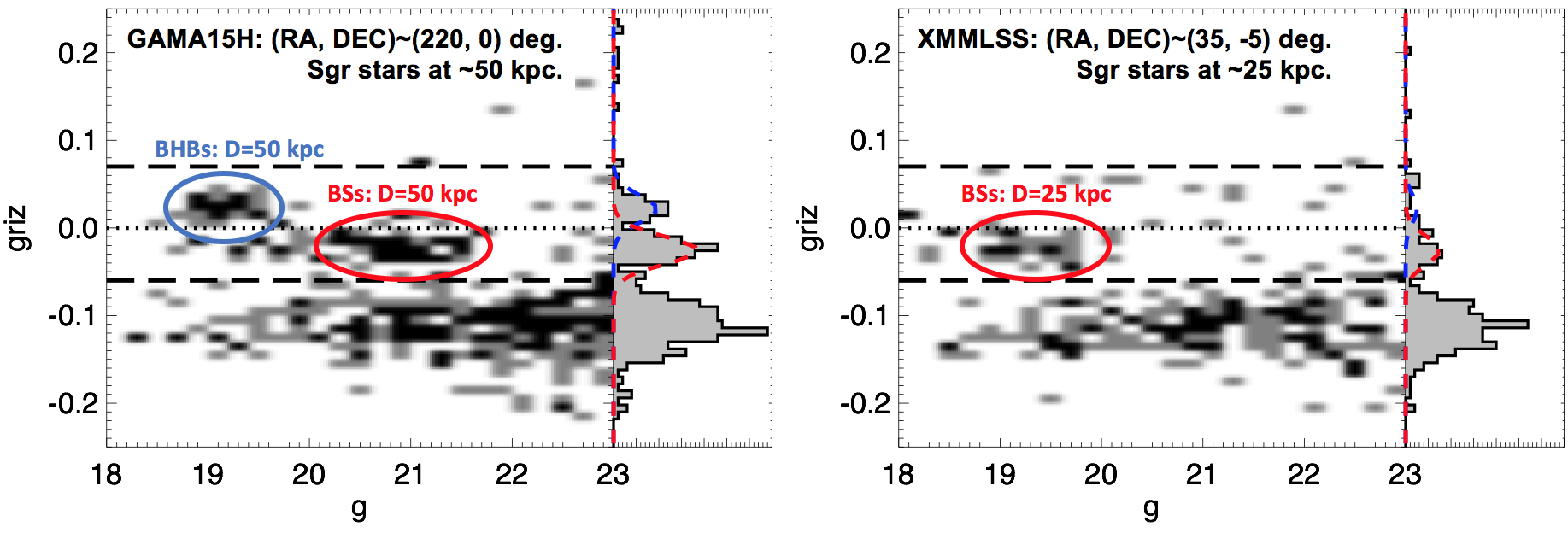}
    \caption{The $griz$ color against $g$-band magnitude for the two HSC fields that overlap with the Sgr stream (left panel = GAMA15H , right panel = XMMLSS). At the locations of these fields, Sgr stars are present at $D\sim 50$ kpc and $D \sim 25$ kpc, respectively. The density plot is given in pixel sizes of 0.2 $\times$ 0.01, and the shading is saturated at $N=2$ stars per pixel. The inset panels show histograms of the $griz$ distribution, where approximate Gaussian fits for the BHB and BS populations are indicated with the dashed blue and red lines, respectively.}
    \label{fig:griz_sgr}
\end{figure*}

To further illustrate the $griz$ decomposition of the BHB and BS stars we show $griz$ against $g$-band magnitude for the two HSC fields that overlap with the Sgr stream (GAMA15H and XMMLSS, see Fig. \ref{fig:fields}) in Fig. \ref{fig:griz_sgr}. At the locations of these fields, Sgr stars are present at $D\sim 50$ kpc and $D \sim 25$ kpc, respectively (see Figure 7 in \citealt{deason12}). The plot shows that there are indeed overdensities of BHB and BS stars in the magnitude ranges corresponding to these distances (assuming $M_g(\rm BHB) \sim 0.5$ and $M_g(\rm BS) \sim 2.5$). Moreover, the excess of BHB and BS stars associated with Sgr occupy narrow sequences in $griz$ space. The right-hand inset panels show histograms of the $griz$ distribution where approximate Gaussian fits for the BHB and BS populations are indicated with the dashed blue and red lines, respectively. Here, the centers and intrinsic widths are fixed according to the values given in Table \ref{tab:cent_widths}. Thus, the $griz$ decomposition allows us to separate BHB and BS stars relatively cleanly, even at faint magnitudes (down to $g \sim 22$).

\subsection{Maximum Likelihood Analysis}
We count the number of BHB stars in bins of 0.5 mag in distance modulus between $18.5 < g-M_g(\rm BHB) < 22$. This corresponds to distances between 50 kpc and 200 kpc. For each distance modulus bin, the probability distribution function is given by:
\begin{eqnarray}
P(\mathbf{x}) &= &f_{\rm BHB}P(\mathbf{x}|\mathrm{BHB}) +f_{\rm BS} P(\mathbf{x} | \mathrm{BS})+\\ \notag
&&f_{\rm WD}\left[0.7 P(\mathbf{x} | \mathrm{WD_{DA}}) +0.3P(\mathbf{x} | \mathrm{WD_{DB}})\right] +\\ \notag
&&\frac{f_{\rm QSO}}{\textbf{x}_{\rm max}-\textbf{x}_{\rm min}}
\end{eqnarray}
where $P(\mathbf{x} | \rm type) = \frac{1}{\sqrt{2\pi}\sigma} \mathrm{exp}\left(-(x-x_0)^2 / 2\sigma^2\right)$ and $\mathbf{x}=griz$. 
Here, the centres and widths of the Gaussians are fixed and only the amplitudes are free parameters. Note that we also include a contribution from QSOs (constant in $griz$) for any residual contaminants. Thus, the number of BHBs in a given distance modulus bin is given by: $N_{\rm BHB} = N_{\rm tot} \times f_{\rm BHB}$. 
The log-likelihood function is constructed from the density distribution:
\begin{equation}
\mathrm{log} \mathcal{L} = \sum^{N_{\rm tot}}_i \mathrm{log} P \left( \mathbf{x}_i \right)
\end{equation}
The log-likelihood is maximized to find the best-fitting $N_{\rm BHB}$, $N_{\rm BS}$ and $N_{\rm WD}$ parameters\footnote{Note $N_{\rm QSO} = N_{\rm tot} - (N_{\rm BHB} + N_{\rm BS} + N_{\rm WD})$} using a brute-force grid search. For each bin in BHB distance modulus, we have an estimate of the number of BHB stars. Thus, we can convert this number count to stellar number density to compute the density profile out to 200 kpc. We estimate the errors in $N_{\rm BHB}$ by marginalizing the likelihood distribution over $N_{\rm BS}$ and $N_{\rm WD}$ and computing the 68\% confidence interval of the 1D marginalized likelihood distribution. Note that these errors are larger than a simple Poisson noise ($\sim \sqrt{N_{\rm BHB}}$) estimate because the distributions are generally non-Poisson (and become more non-Poisson with greater overlap between the BHB, BS and WD populations).

\section{BHB Star Number Counts}
\label{sec:counts}

\begin{figure*}
       \centering
        \includegraphics[width=16cm, height=8cm]{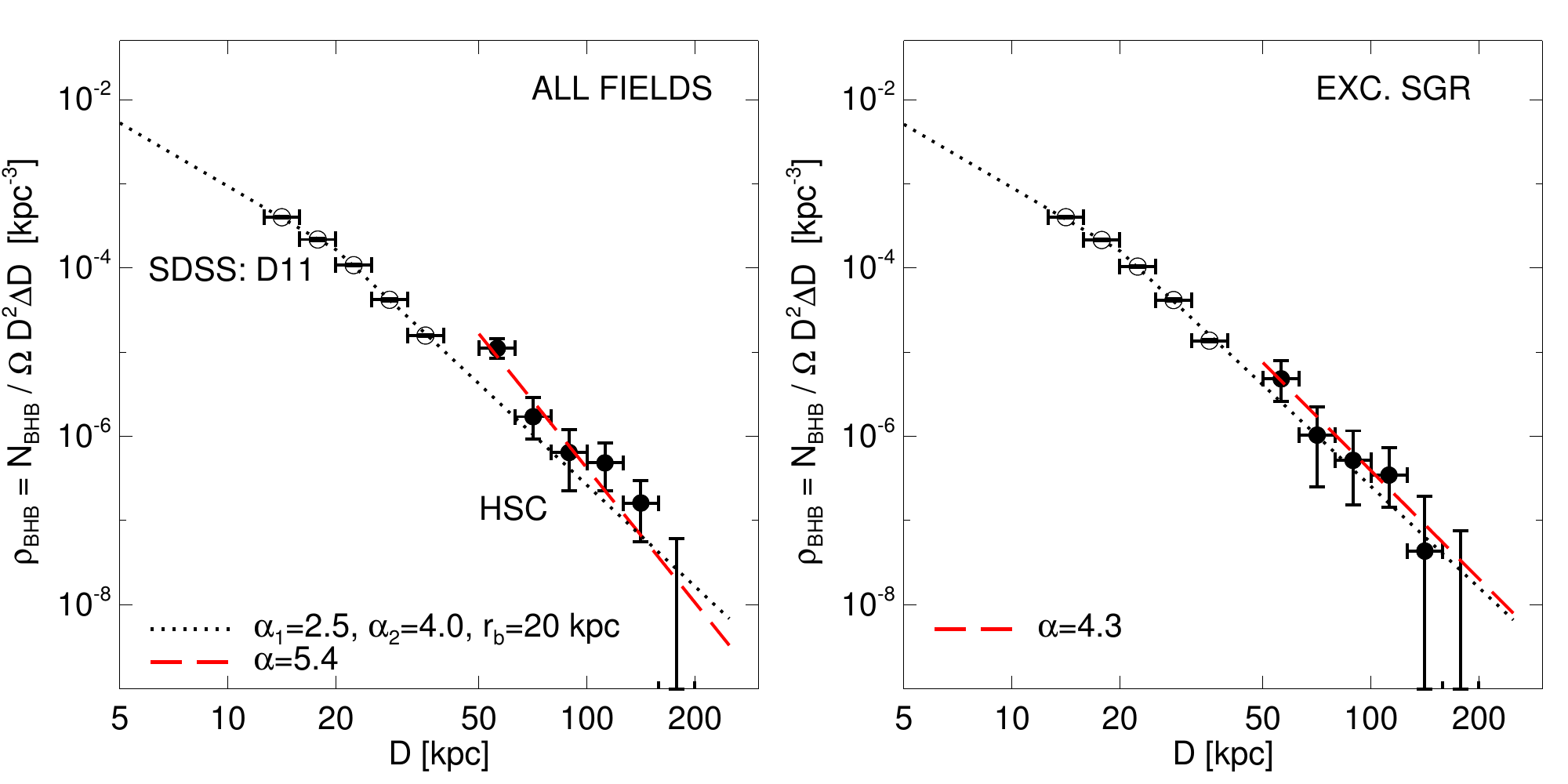}
        \caption{The density profile of BHB  stars as a function of Heliocentric distance. The left-hand panels include all Hyper Suprime-Cam Wide fields, and the fields overlapping with the Sgr stream are removed in the right-hand panels. The solid black points indicate the results from the HSC data. The red dashed lines indicate a single-power law fit. The open squares give the approximate densities of BHB stars from SDSS at smaller distances from \cite{deason11}. The dotted line shows an extrapolation of the SDSS results to larger distances. The influence of Sgr is clear at $\sim 50$ kpc, here there is an excess of BHB stars followed by a rapid decline beyond the Sgr apocentre. When Sgr is removed, the counts in HSC are consistent with a continuation of a $\alpha \sim 4$ power-law from smaller distances.}
    \label{fig:rho_bhb}
\end{figure*}

\begin{figure}
        \includegraphics[width=8.5cm, height=4.25cm]{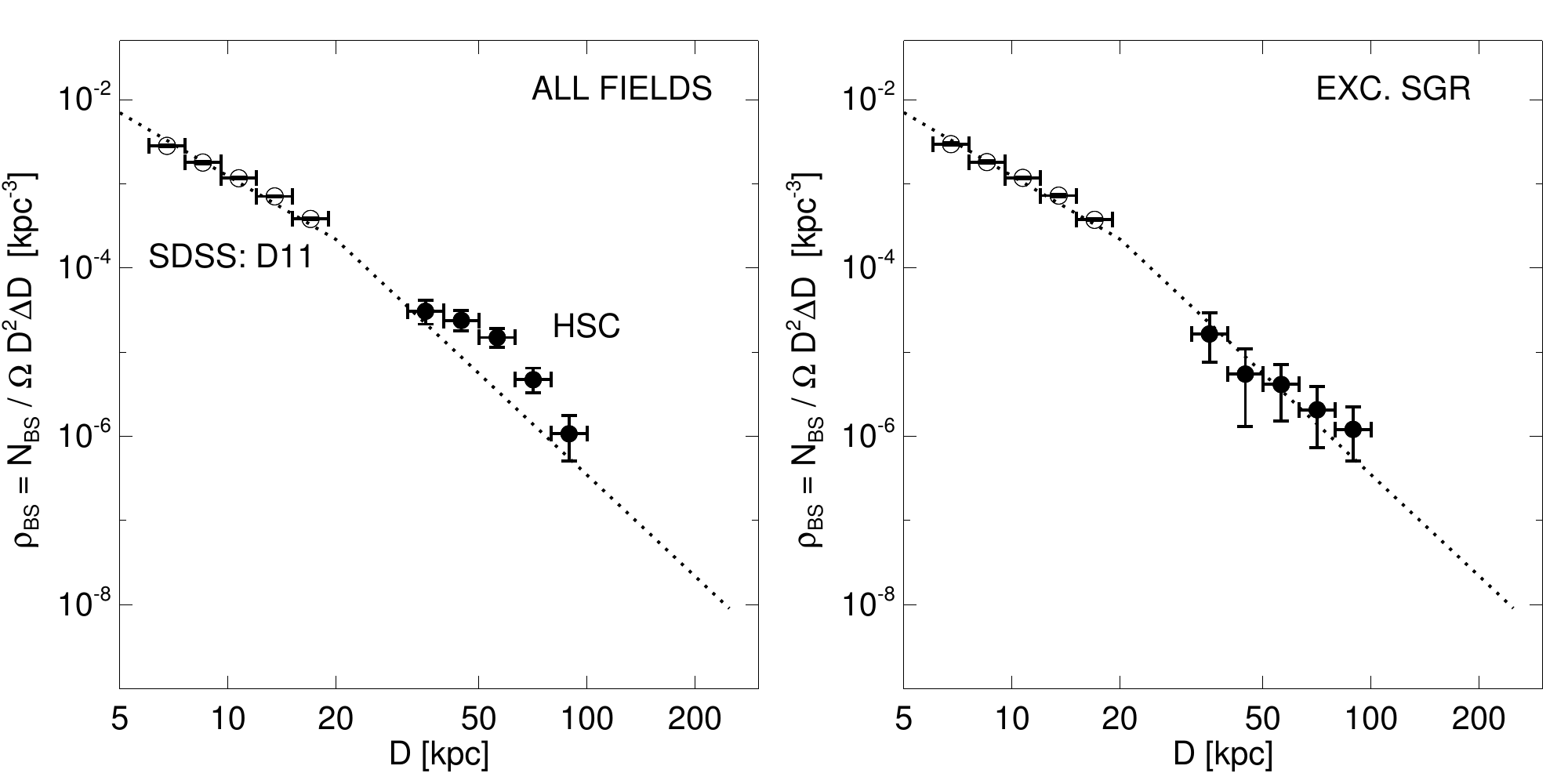}
        \caption{As Fig. \ref{fig:rho_bhb}, but for BS stars.}
    \label{fig:rho_bs}
\end{figure}

\begin{figure}
     \includegraphics[width=8cm, height=6.4cm]{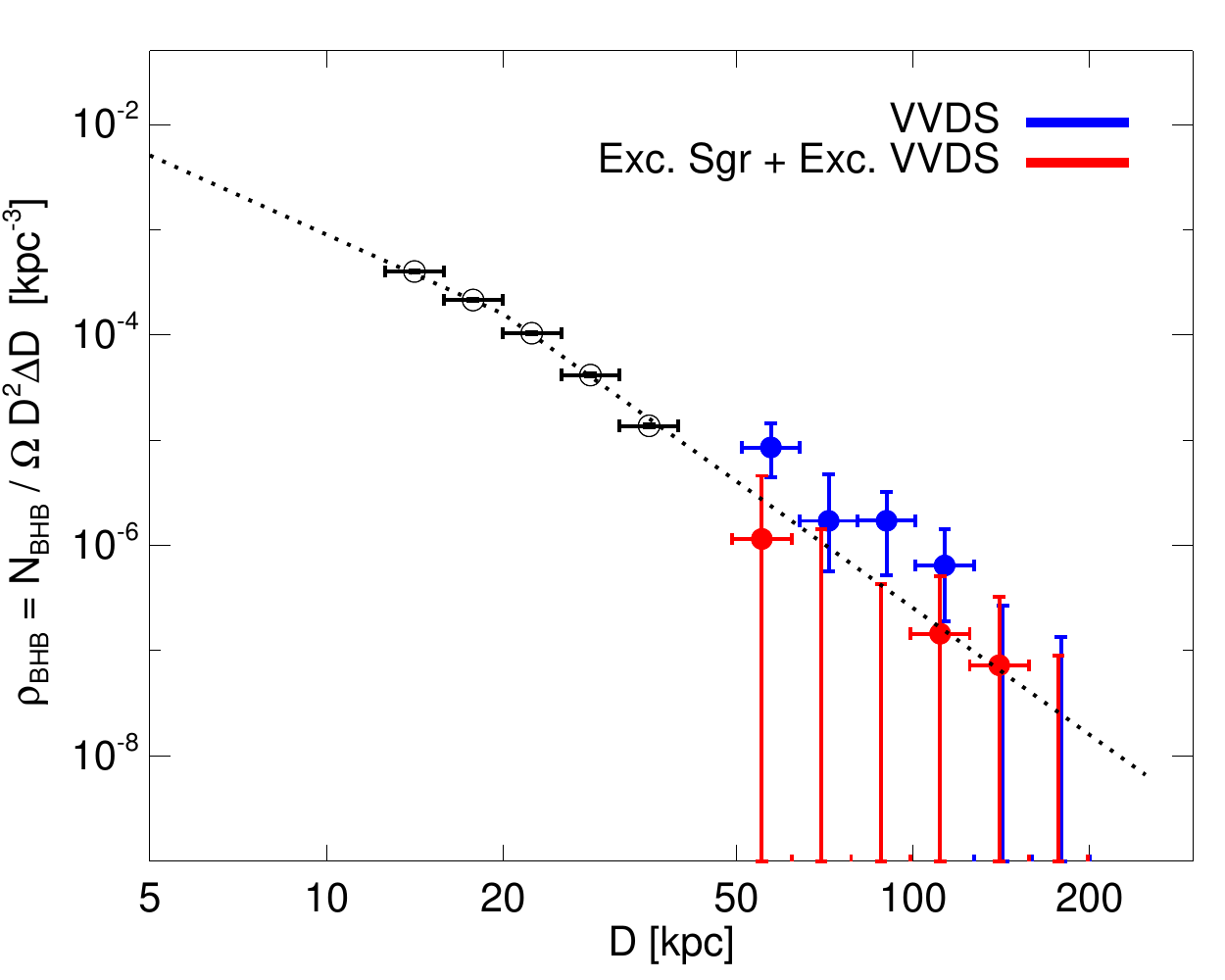}
     \caption{The density profile of BHB stars for fields excluding the Sgr stream. Here, we show the counts for the VVDS field in blue and the remaining (non Sgr) fields in red. There is a notable excess of BHB stars in the VVDS field. In fact, almost all of the BHB stars in the non-Sgr fields are found in VVDS.}
    \label{fig:rho_fields}
\end{figure}

\begin{figure*}
    \centering
    \includegraphics[width=17.5cm, height=9.55cm]{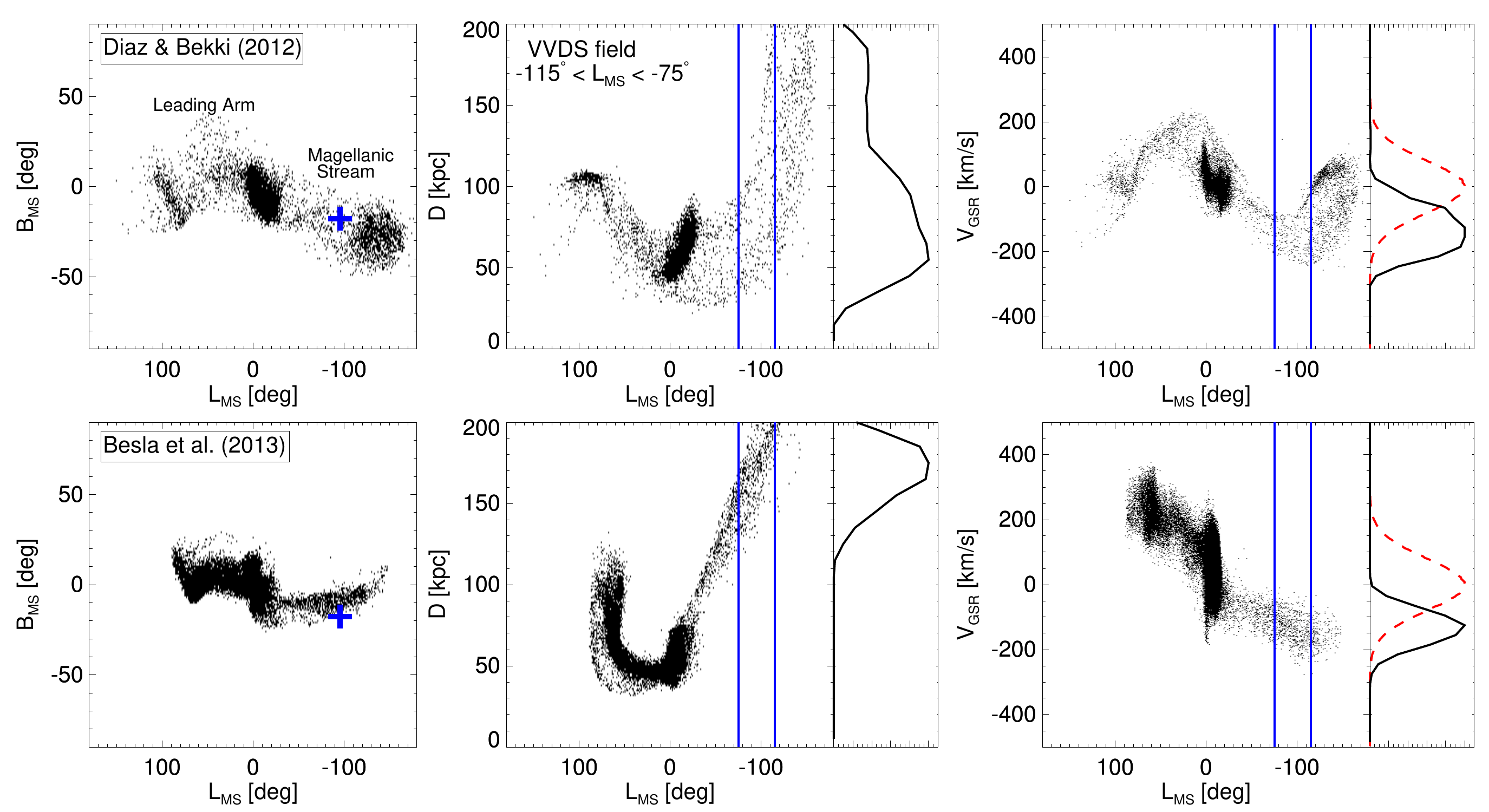}
    \caption{Models of Magellanic debris. Models from \cite{Diaz2012} and \cite{besla13} are shown in the top and bottom panels, respectively. \textit{Left panels:} The distribution of debris on the sky in Magellanic coordinates. The blue cross indicates the position of the VVDS field. The location of the Leading Arm and Magellanic Stream are indicated. \textit{Middle panels:} Distance against Magellanic longitude.  In the region of the VVDS field ($-115^\circ < L_{\rm MS} < -75^\circ$) there is potential debris in the distance range $50 < D/\mathrm{kpc} < 200$. The inset panel shows a histogram of distances in the  $-115^\circ < L_{\rm MS} < -75^\circ$ slice. \textit{Right panels:} Galactocentric velocity against Magellanic longitude. The black line in the inset panels shows the distribution of velocities in the region of the VVDS field, and the dashed red line indicates a halo population with $\sigma = 80$ km s$^{-1}$. Line-of-sight velocities could be used to identify Magellanic debris in the VVDS field.}
    \label{fig:mag_sims}
\end{figure*}

In Fig. \ref{fig:rho_bhb} we show the density of BHB stars as a function of heliocentric distance. The solid points show our results from the HSC data. In the left panels, all HSC fields are used. In the right panels, those fields that overlap with the Sgr stream are excluded (see Fig. \ref{fig:fields}). The open circles give the approximate number counts of BHB stars from SDSS for $D < 40$ kpc derived from  \cite{deason11}. The dotted line shows a ``broken'' density profile with $\alpha_1=2.5, \alpha_2=4.0$ and $r_b=25$ kpc, which is a good-fit to the inner stellar halo density profile. We also fit power-law density profile models to the HSC data between $50 < D/\mathrm{kpc} < 200$. Here, we use a $\chi^2$ routine to find the best-fitting parameter values. The dashed red line shows the best-fit single power-law profile ($\rho \propto r^{-\alpha}$) with $\alpha=5.4 \pm 1.4$. When we exclude fields overlapping with the Sgr stream, a power-law of $\alpha=4.3 \pm 1.2 $ is a good fit to the data, and the HSC counts are consistent with the continuation of an $\alpha \sim 4.0$ power-law from smaller distances.

The influence of the Sgr stream is clear at D $\sim 50$ kpc. This distance approximately corresponds to the apocentre of the of the Sgr leading arm (see Figure 4 of \citealt{belokurov14}). Here, there is an excess of BHB stars followed by a rapid decline just beyond the Sgr apocentre (cf. \citealt{deason14}). However, when the Sgr fields are excluded from the analysis, the counts in HSC are consistent with a continuation of a $\alpha \sim 4$ power-law from smaller distances.

In Fig. \ref{fig:rho_bs} we show the density of BS stars. Here we use the same procedure as for BHBs, and adopt the BS absolute magnitude calibration from \cite{deason11}. Note that the absolute magnitude calibration for BS stars is more uncertain that BHBs, with $\sigma(M_{\rm BS}) \sim 0.5$ (see \cite{deason11} Figure 4). Thus, we show the BS density profile for illustration only. Nonetheless, the influence of Sgr is also clear in the BS number counts, where there is an excess of stars near $D \sim 50$ kpc followed by a sharp drop-off. Again, when the Sgr fields are excluded, the HSC BS star counts are consistent with an $\alpha=4$ power-law.

We explore the field-to-field variation in the HSC photometry in Fig. \ref{fig:rho_fields}. Here, we show the BHB density profile for fields excluding Sgr. The counts for the VVDS field are shown in blue, and the remaining (non-Sgr) fields are shown in red (see Fig. \ref{fig:fields}).  The VVDS field has a notable excess of BHB stars. In fact, almost all of the BHB stars in the non-Sgr fields are found in VVDS. This apparent excess in VVDS is intriguing, and may, of course, simply be due to field-to-field variation. However, the apparent excess cannot be explained by pure Poisson fluctuation; the probability of observing 17 or more stars in 27 deg$^2$ of VVDS given the 3 stars observed in the other (non-Sgr) 40 deg$^2$ of HSC is less than $1.2 \times 10^{-5}$. Note that for this calculation, we assume that the the 40 deg$^2$ of non-Sgr coverage (excluding VVDS) represents the ``field'' halo distribution. We also investigate the number counts of BS stars in the VVDS field. We find that there is indeed an excess of BS stars in this field between $\sim 50-100$ kpc ($N=14$ in 27 deg$^2$ of VVDS, $N=7$ in the remaining 40 deg$^2$), which provides further evidence that the VVDS field overlaps with substructure in this radial range.

The HSC data release 1 paper \citep{aihara17} describes how a small number of patches in the VVDS field suffer from a PSF modeling problem. In these cases, the PSF could not be accurately modeled due to extremely good seeing. However, we find that most of our candidate BHBs in the VVDS field do not overlap with these ``bad patches'', and we find no evidence that the apparent excess is caused by the poor photometry in parts of the field. For example, if we cull the regions with PSF modeling problems, we are still left with 14 BHB stars in the VVDS field. Thus, we conclude that the excess of BHB stars is not due to statistical noise or observing variations between fields, and is most likely caused by substructure in the stellar halo.

In Fig. \ref{fig:fields} we saw that models of Magellanic Cloud disruption can predict the presence of stellar debris in the vicinity of the VVDS field. Below, we describe these Magellanic debris models and explore the possibility that the excess in VVDS could have a Magellanic origin.

\section{Magellanic debris beyond 50 kpc?}
\label{sec:magellanic}

In Fig. \ref{fig:mag_sims} we show the predicted debris from two models of Magellanic cloud disruption. In the top panels we show a model chosen to reproduce the results in \cite{Diaz2012}. This model follows the disruption of the SMC in the presence of the LMC using the Lagrange cloud stripping technique of \cite{gibbons14}. The simulation follows a similar setup to \cite{Diaz2012} and was chosen to reproduce the H\textsc{i} features in the Magellanic Stream and Bridge. Here, the LMC and SMC have masses of $1 \times 10^{10} M_\odot$ and $3 \times 10^9 M_\odot$, and are simulated in a three-component MW potential made up of a Miyamoto-Nagai disk, a Hernquist bulge, and an NFW halo (see \citealt{deason17, belokurov17} for more details). The bottom panels show a model from \cite{besla13} generated using the smoothed particle hydrodynamics code \textsc{gadget}-3 \citep{springel05}. Here, the LMC has a total mass of $1.8 \times 10^{11} M_\odot$, the SMC has a total mass of $2 \times 10^{10} M_\odot$, and the MW is modeled as a static NFW potential with a total mass of $1.5 \times 10^{12} M_\odot$. In \cite{besla13} two models are generated with this setup. These models differ in terms of the orbital interaction history of the Clouds; we show the Model 2, which shows better agreement with the internal structure and kinematics of the LMC (see \citealt{besla12}.)

The model debris is shown in Magellanic coordinates in the left-hand panel of Fig. \ref{fig:mag_sims}. The location of the VVDS field is indicated with the blue cross; the field lies close to the Magellanic plane ($B_{MS} \sim -18^\circ$). The middle panels show the distance distribution of the debris. In both models,  Magellanic debris is predicted to reach from $\sim 50$ kpc to 200 kpc at the position of the VVDS. Note that there is no constraint on these distances in the modeling, and the exact details of the modeling procedure can significantly change the distance range of debris at the position of the VVDS. Nonetheless, it is clear that the distance range of our BHB excess approximately overlaps with the predicted range of stellar debris in the Magellanic stream. Moreover, confirmation of Magellanic debris in this region of the sky, and the distances of the BHB stars, could significantly constrain models for the recent interaction history of the Clouds. Finally, we show the predicted line-of-sight velocity distribution of the Magellanic debris in the right-hand panel of Fig. \ref{fig:mag_sims}. We show the approximate velocity distribution of halo stars with the dashed red line (assuming $\sigma_{\rm los} = 80$ km s$^{-1}$, see \citealt{deason12}). Here, the velocities in the region of the VVDS are offset from the ``field'' stellar halo population. Thus, follow-up spectroscopy of the BHB stars that we have identified in VVDS could confirm the presence of Magellanic debris out to such large distances in the Galactic halo.

We have found an intriguing excess of BHB stars in the VVDS field in the distance range $50 < D/\mathrm{kpc} < 200$. While this excess could be due to field-to-field variation, it is plausible that we have detected Magellanic debris out to significant distances in the halo. With future data-releases of HSC photometry covering a wider area of the sky, and spectroscopic follow-up of the BHB excess, we should be able to confirm or falsify this scenario. Moreover, if indeed we have detected a notable substructure at large distances, our results suggests that the density of ``field'' BHB stars beyond 50 kpc are very low, and consistent with a rapid fall off (power-law with index $\alpha \ge 4$) in counts beyond 50 kpc.

\section{Summary and Discussion}
\label{sec:final}
We have used deep HSC photometry to measure the density profile of BHB stars beyond 50 kpc in the Galactic halo. Using the $\sim 100$ deg$^2$ HSC wide fields, we identify BHB stars using multi-band $g, r, i, z$ photometry. The combination of filters, $griz =i-z-0.3(g-r)+0.035 $ shows a distinct sequence of BHB stars, which we use to  model the number counts of BHB stars as a function of magnitude and allow for contributions from BS, WD and QSO contaminants.

Our results are consistent with a continuation of a $-4$ power-law from the inner halo. However, we find that almost all of the distant BHB stars could be associated with known, massive substructures; namely, the Sgr stream and the Magellanic Clouds. The extent of the Sgr stream is well-known (e.g. \citealt{belokurov14, Sesar2017}), and we indeed see a notable excess of stars near the apocentre of the Leading arm ($\sim 50-60$ kpc). For the non-Sgr HSC fields, almost all of the distant BHB stars are associated with one field called VVDS, which lies close to the Magellanic plane. Comparison with models of Magellanic cloud disruption shows that the location and extent of these distant BHB stars are consistent with being tidal debris in the outer reaches of the Galaxy. The existence of stellar debris associated with the Magellanic Stream is yet to be unambiguously confirmed, and our results could herald the first detection of Magellanic stars out to the virial radius of the Galaxy.

If the stars in VVDS are indeed Magellanic debris or an as-yet unidentified substructure, our work raises an important question: Where are the rest of the halo stars beyond 50 kpc? The outer realms of the stellar halo are predicted to be lumpy and un-mixed, but there are few detections of significant number counts of halo stars beyond this distance. Indeed, our results suggest that the outer halo, which presumably probes late-time accretion events, is dominated by the small number of recent Milky Way digestions (i.e. Sgr and the Clouds): The remaining ``field'' halo stars at large distances are few and far between, perhaps because, quite simply, the Galaxy has had its fill. 
\\
\\
\noindent
In this work we exploited the first $\sim 100$ deg$^2$ of the HSC Wide imaging survey. At completion, this survey will cover an area over 10 times larger ($1,400$ deg$^2$) to similar depth. Thus, this work is a mere taster for the future HSC data releases that we can use to probe the very distant halo. Looking further ahead, the HSC precedes the Large Synoptic Survey telescope, which will cover half of the sky in $u, g, r, i, z$ down to $r \sim 27$. This unprecedented survey will surely revolutionize our understanding of the outer halo, and, quite possibly, push us beyond the perceived periphery of the Galaxy.

 \acknowledgments
We thank Gurtina Besla for providing her simulation data and for useful comments on this work. We also thank an anonymous referee for providing constructive comments on our paper. A.D thanks S.O.C for timing her arrival to perfection!

A.D. is supported by a Royal Society University Research Fellowship. 
A.D. also acknowledges support from the STFC grant ST/P000451/1.
The research leading to these results has received funding from the
European Research Council under the European Union's Seventh Framework
Programme (FP/2007-2013) / ERC Grant Agreement n. 308024. V.B. acknowledges financial support from the ERC. 

The Hyper Suprime-Cam (HSC) collaboration includes the astronomical communities of Japan and Taiwan, and Princeton University. The HSC instrumentation and software were developed by the National Astronomical Observatory of Japan (NAOJ), the Kavli Institute for the Physics and Mathematics of the Universe (Kavli IPMU), the University of Tokyo, the High Energy Accelerator Research Organization (KEK), the Academia Sinica Institute for Astronomy and Astrophysics in Taiwan (ASIAA), and Princeton University. Funding was contributed by the FIRST program from Japanese Cabinet Office, the Ministry of Education, Culture, Sports, Science and Technology (MEXT), the Japan Society for the Promotion of Science (JSPS), Japan Science and Technology Agency (JST), the Toray Science Foundation, NAOJ, Kavli IPMU, KEK, ASIAA, and Princeton University. 

This paper makes use of software developed for the Large Synoptic Survey Telescope. We thank the LSST Project for making their code available as free software at  \url{http://dm.lsst.org}.

The Pan-STARRS1 Surveys (PS1) have been made possible through contributions of the Institute for Astronomy, the University of Hawaii, the Pan-STARRS Project Office, the Max-Planck Society and its participating institutes, the Max Planck Institute for Astronomy, Heidelberg and the Max Planck Institute for Extraterrestrial Physics, Garching, The Johns Hopkins University, Durham University, the University of Edinburgh, Queen?s University Belfast, the Harvard-Smithsonian Center for Astrophysics, the Las Cumbres Observatory Global Telescope Network Incorporated, the National Central University of Taiwan, the Space Telescope Science Institute, the National Aeronautics and Space Administration under Grant No. NNX08AR22G issued through the Planetary Science Division of the NASA Science Mission Directorate, the National Science Foundation under Grant No. AST-1238877, the University of Maryland, and Eotvos Lorand University (ELTE) and the Los Alamos National Laboratory.

Based on data collected at the Subaru Telescope and retrieved from the HSC data archive system, which is operated by Subaru Telescope and Astronomy Data Center at National Astronomical Observatory of Japan.

\bibliographystyle{aasjournal}
\bibliography{mybib}

\end{document}